\documentclass[twocolumn,showpacs,superscriptaddress,amsmath,amssymb,pra]{revtex4}
\usepackage{graphicx}
\begin{document}

\title{A supercircle description of universal three-body states in two dimensions}
\author{F.~F. Bellotti}
\affiliation{Instituto Tecnol\'{o}gico de Aeron\'autica, 12228-900, S\~ao Jos\'e dos Campos, SP, Brazil}
\affiliation{Instituto de Fomento e Coordena\c{c}{\~a}o Industrial, 12228-901, S{\~a}o Jos{\'e} dos Campos, SP, Brazil}
\author{T. Frederico}
\affiliation{Instituto Tecnol\'{o}gico de Aeron\'autica, 12228-900, S\~ao Jos\'e dos Campos, SP, Brazil} 
\author{M.~T. Yamashita}
\affiliation{Instituto de F\'\i sica Te\'orica, UNESP - Univ Estadual Paulista, C.P. 70532-2, CEP 01156-970, S\~ao Paulo, SP, Brazil} 
\author{D.~V. Fedorov}
\author{A.~S. Jensen}
\author{N.~T. Zinner}
\affiliation{Department of Physics and Astronomy, Aarhus University, DK-8000 Aarhus C, Denmark}
\date{\today }

\begin{abstract}
Bound states of asymmetric three-body systems
confined to two dimensions are currently unknown.
In the universal regime, 
two energy ratios and two mass ratios
provide complete knowledge of the three-body energy measured in units
of one two-body energy. We compute the three-body energy 
for general systems using numerical momentum-space
techniques.
The lowest number of stable bound states is
produced when one mass is larger than two similar masses. We
focus on selected asymmetric systems of interest in cold atom
physics.  The scaled three-body energy and the two scaled two-body
energies are related through an equation for a {\it supercircle} whose
radius increases almost linearly with three-body energy.  The
exponents exhibit an increasing behavior with three-body
energy. The mass dependence is highly non-trivial. 
Based on our numerical findings, we give a simple
relation that predicts the universal three-body energy.
\end{abstract}
\maketitle

\section{Introduction.}
The spatial dimensions are crucial for the behavior of quantum systems.
The reason for this can be understood by considering the kinetic energy
only, and observing that it behaves vastly different in two dimensions
as compared to one and three. In fact, the presence of a negative 
centrifugal barrier for zero angular momentum 
in two dimensions implies that an infinitesimal 
attractive potential produces a {\it universal} bound state \cite{landau,vol11}.
This interesting non-trivial role of the spatial dimensionality can
be isolated and studied in detail in the so-called universal limit
where the effect of interactions can be described by one or two
{\it model-independent} quantities and the detailed structure of 
two-body potentials at short-range thus becomes irrelevant. 

The implementation of Feshbach resonances in cold atomic
gases \cite{chin2010} allows experimenters to tune their systems
into the universal regime and study properties that are independent
of the particularities of the atoms and molecules. One of the 
great successes of this program is the observation of three-body
Efimov states \cite{kraemer2006,ferlaino2010} which had been 
proposed but never found in nuclear physics \cite{jen04}. The
presence of resonances also allows the study of the strongly 
correlated unitary limit, where the physics is universal and
governed by a single so-called contact parameter \cite{tan2008} that
is measurable \cite{stewart2010,kuhnle2010}.
Currently there is an experimental push to reduce the ultracold 
systems from three to two dimensions and study the universal 
behavior \cite{turlapov2010,frohlich2011,dyke2011}. The highly 
desired theoretical knowledge of strongly correlated 
gases is more sparse than in the three-dimensional case. 
Recently, the contact has been derived using different 
approaches \cite{combescot2009,castin2010,valiente2011,braaten2011}.
The question of few-body states in equal mass 
two-dimensional systems has been addressed in a number of works.
Bound states of three- \cite{tjo75,adh88,nie97,nie01,brod2006,kart2006}, 
four- \cite{platter2004,brod2006}, and larger particle 
number droplets \cite{hammer2004,blume2005,lee2006}, and 
scattering \cite{brod2006} and recombination observables \cite{helfrich2011}
have been addressed.

 The general three-body bound state spectrum in the universal limit in two
dimensions has not previously been investigated. The limiting case of
three identical bosons is known to have exactly one two-body and two
three-body bound states in the universal limit
\cite{tjo75,adh88,nie97,nie01}, and all ratios of energies and radii
are universal constants.  
This is in marked contrast to three dimensions where the Efimov
effect produces an infinite ladder of states when the scattering
length diverges \cite{efi70}. While the results in three dimensions
can be obtained through analytical means, the spectrum in 
the two-dimensional case has been obtained through numerical
methods only,
and no analytical insights have been found so far. Furthermore, 
the problem of three particles with different masses
and short-range interactions in two dimensions has not been 
addressed before. The combination of 
three different species of atoms was recently reported \cite{wu2011}. 
Although that experiment is presently in three dimensions, it 
should be easily extended to a two-dimensional setup.

The universal limit can be considered in a natural way through 
zero-range interactions that leave only a single parameter 
in the potential to be related to the scattering length or 
two-body bound state energy. 
Here we numerically solve the momentum-space integral equations in two dimensions
\cite{bel11}. In three dimensions similar formalism has been 
discussed within the language of effective field theory \cite{braaten2006}.
We extract universal features of completely asymmetric three-body
systems and provide simple parametrizations of energies as functions
of the two-body properties in terms of Lam{\'e} curves \cite{lame1818}. 
This provides a uniform parametrization of 
all aspects of the three-body bound state spectrum in two dimensions
that can aid the search for a (semi-)analytical understanding of 
these intruiging systems.

\begin{figure}[ht!]
\includegraphics[scale=0.28,clip=true]{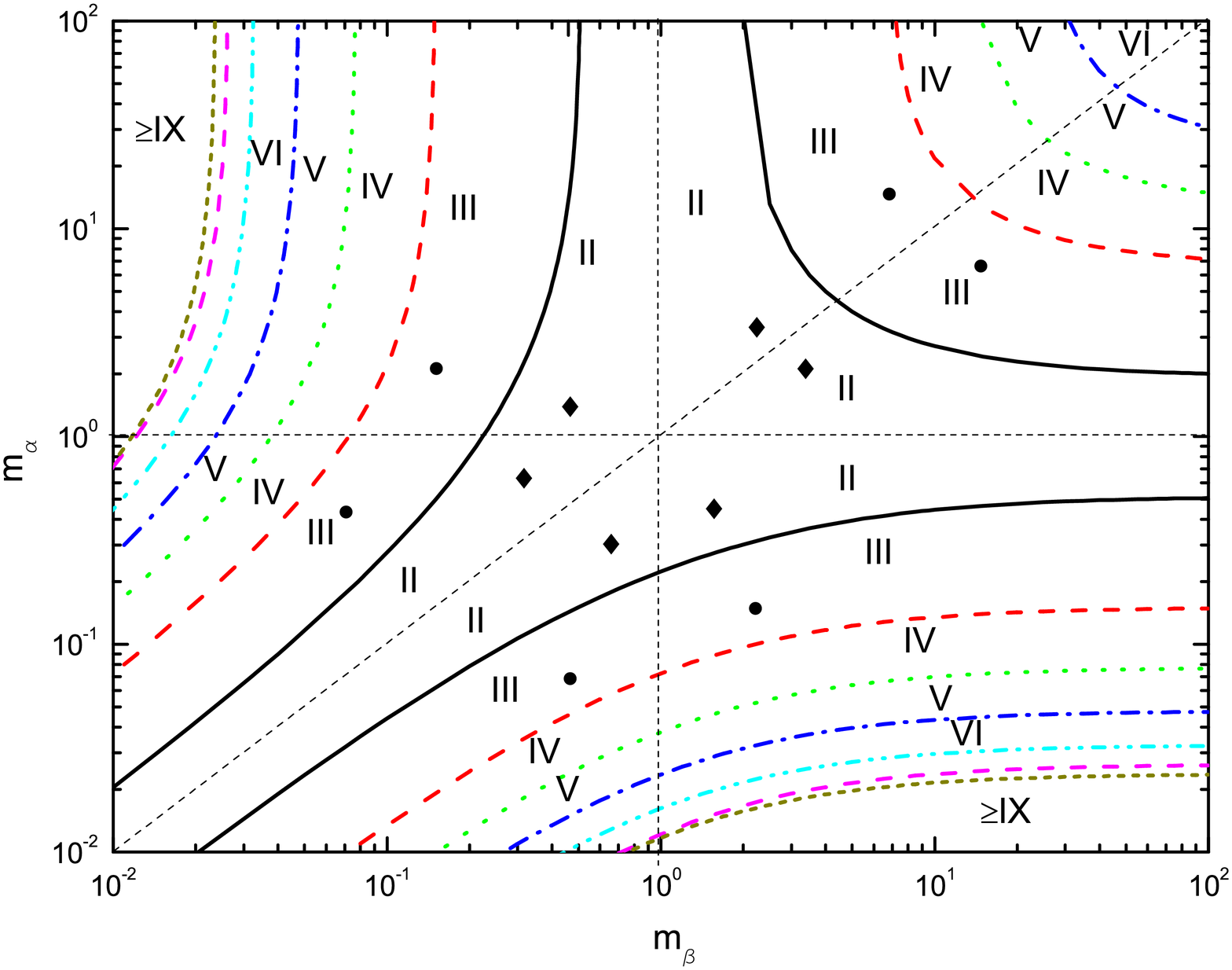}
\caption{Phase diagram of the maximum number of three-body bound states as
  functions of two mass ratios, $m_{\alpha}$ and $m_{\beta}$.  The
  three two-body energies are equal, $E_{ab}=E_{bc}=E_{ac}$.
  The roman numerals indicated the maximum number of bound states in each region. The
  circular ($^6$Li$^{40}$K$^{87}$Rb) and diamond ($^{40}$K$^{87}$Rb$^{133}$Cs) points are the realistic cases studied in
  Figs.~\ref{fig3} and \ref{fig4}. The two sets of points are related by the symmetries in Eq.~\eqref{e57}.}
\label{fig1}
\end{figure}

\section{Procedure and definitions.}
We consider three particles, $a$, $b$, and $c$ of mass $m_a$, $m_b$, and $m_c$ 
with pairwise contact interactions which produce two-body bound states of
energies $E_{ab}$, $E_{ac}$, and $E_{bc}$.  The wave equation
in momentum space is a set of three coupled integro-differential
equations \cite{tjo75,bel11,bel+11}. 
Here we provide a full survey of the three-body energies,
$E_3$, in the most general setup. Initially $E_3$ 
is a function of six parameters.  However,
the use of two-body energies instead of interaction strengths implies
that only mass and energy ratios enter the equations.  This means that
$\epsilon_3=E_3/E_{ab}$ divided by one of the two-body energies can be expressed as a
function of four dimensionless parameters, i.e.
\begin{align}
 \epsilon_3 = F_n\left(\frac{E_{bc}}{E_{ab}},
\frac{E_{ac}}{E_{ab}}, \frac{m_b}{m_a}, \frac{m_c}{m_a} \right)
 \equiv F_n(\epsilon_{bc},\epsilon_{ac},m_{\alpha},m_{\beta}),
\label{e40}
\end{align}
where $\epsilon_3$ is the scaled
three-body energy, $m_{\alpha}=m_b/m_a$, $m_\beta=m_c/m_a$, 
$\epsilon_{bc}=E_{bc}/E_{ab}$, and $\epsilon_{ac}=E_{ac}/E_{ab}$.
The universal functions, $F_n$, are labeled by the subscript $n$ to
distinguish between ground, $n=0$, and excited states, $n>0$.
By interchange of particle labels we find that all the
universal functions, $F_n$, must obey the symmetry relations:
\begin{align} \label{e57}
& F_n\left(\epsilon_{bc},\epsilon_{ac},m_{\alpha},m_{\beta}\right) = 
 F_n\left(\epsilon_{ac},\epsilon_{bc},m_{\beta},m_{\alpha}\right)  =& \\ \nonumber &
 \epsilon_{bc} 
 F_n\left(\frac{1}{\epsilon_{bc}},
 \frac{\epsilon_{ac}}{\epsilon_{bc}},
 \frac{1}{m_{\alpha}},\frac{m_{\beta}}{m_{\alpha}}\right) =
 \epsilon_{bc}
 F_n\left(\frac{\epsilon_{ac}}{\epsilon_{bc}},\frac{1}{\epsilon_{bc}},
\frac{m_{\beta}}{m_{\alpha}},
 \frac{1}{m_{\alpha}}\right)=& \\  \nonumber  & \epsilon_{ac}
F_n\left(\frac{1}{\epsilon_{ac}},\frac{\epsilon_{bc}}{\epsilon_{ac}},
\frac{1}{m_{\beta}},\frac{m_{\alpha}}{m_{\beta}}\right)
 = \epsilon_{ac}
 F_n\left(\frac{\epsilon_{bc}}{\epsilon_{ac}},\frac{1}{\epsilon_{ac}},
 \frac{m_{\alpha}}{m_{\beta}},\frac{1}{m_{\beta}}\right).&
\end{align}
The energy and
mass scaling leaves us with four non-trivial parameters. Using the
symmetry relations in Eq.~\eqref{e57} we can restrict the
investigations of $F_n$ to smaller regions of this four-parameter
space as indicated by the dashed symmetry lines in Fig.~\ref{fig1}.

\section{Survey of mass dependence.}
The mass dependence of the
number of three-body bound states for a system with $E_{ab}=E_{ac}=E_{bc}$
is shown.
In the central region around equal masses we have
the two bound states \cite{tjo75}. This region,
labeled II, extends in three directions corresponding to one heavy and
two rather similar light particles, that is either $m_{\alpha} \sim
1$, $m_{\beta} \sim 1$, or $m_{\alpha} \simeq m_{\beta} \le 1$.
Moving away from these regions in Fig.~\ref{fig1}, the number of stable
bound states increases in all directions. As an example, consider
$m_\alpha=10$ and vary $m_\beta$ from small to large value in Fig.~\ref{fig1}.
Perhaps surprisingly, along this line
the number of bound states decreases to a minimum of two and 
subsequently increases again. 
The reason is that the disappearing states
merge into the two-body continuum.  The same
behavior is found in three dimensions 
for the disappearance of the infinitely many
Efimov states when the attractive strength is increased \cite{yam02}.
Variation of the two-body energies from all being equal leads to
distortion of the figure but the main structure remains. The central
region still has the smallest number of stable bound states. Any
variation in the two-body energies make region II bigger, pushing the
other lines away from the center. Thus, Fig.~\ref{fig1} shows the
maximum number of stable bound states for one system described by
($m_{\alpha}$, $m_{\beta}$).

\begin{figure}[t!]
\includegraphics[scale=0.28,clip=true]{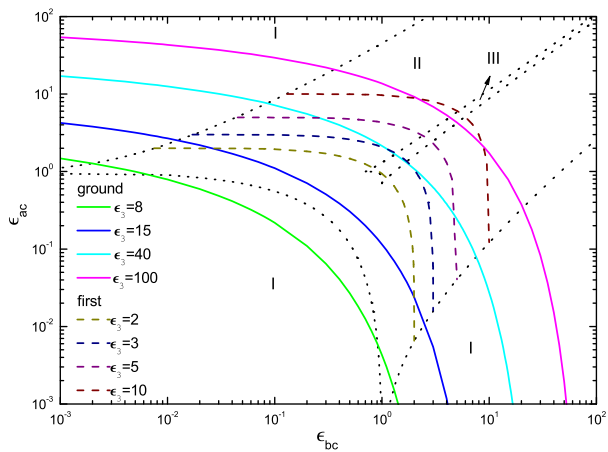}
\caption{Contour diagrams with lines of fixed $\epsilon_3$ values as
  function of the two-body energies $\epsilon_{ac}$ and $\epsilon_{bc}$.
  The solid and dashed curves are for ground and excited states,
  respectively. Here $a$ is $^{87}$Rb, $b$ is $^{40}$K, and $c$ is 
  $^{6}$Li.
  The dotted curves show where the number of stable bound
  states change from one (regions most asymmetric for small energies),
  to two (comparable size of the two-body energies), and to three (equal
  and large energies). The roman labels I, II, and III are as in Fig.~\ref{fig1}. }
\label{fig2}
\end{figure}

\subsection{Three-body energies for given masses.}
Realistic scenarios correspond to given particles (atoms or molecules)
with known masses. In contrast, the interactions are
variable through Feshbach resonances \cite{chin2010}.  
We therefore assume masses corresponding to alkali
atoms $^{87}$Rb, $^{40}$K, and $^{6}$Li. The two ratios of two-body
energies are left as variables where each set uniquely specifies
the three-body energies of ground and possibly excited states.
In Fig.~\ref{fig2}, we show a contour diagram of the scaled three-body
energies for the two lowest stable bound states.  
The log-log plot is very convenient for visibility but can
be deceiving. On a linear scale the curves of equal scaled
three-body energy would look different. However, the parametrizations
considered below nonetheless employ the linear scale.
The chosen set of masses only
allow one, two, or three stable bound states, depending on the two-body
energies. The corresponding regions are shown by dotted curves in
Fig.~\ref{fig2}. The true extent of the regions cannot be seen.  Both
region II and III are closed, e.g region II continues along region III
up to energy ratios of about $10^{\pm5}$, and the narrow region III is
entirely embedded in region II. Other sets of mass ratios could open
region III and allow regions inside with more than three stable bound
states.

In the log-log plot of Fig.~\ref{fig2} we can see the two-body
energies varying by five orders of magnitude, whereas the scaled
three-body energies for a stable system must be larger than all
two-body energies, and in particular larger than unity since it is
measured in units of $E_{ab}$.  The three-body energy contours connect
minimum and maximum two-body energies, that is zero and maximum
two-body energies for the ground state and thresholds boundaries for
existence of the excited states.  The contours appear in regular
intervals with larger values for increasing two-body energies.  Large
three-body energies reflect small spatial extension and therefore are less
interesting as it presumably is unreachable in the universal limit.
The contours pass continuously through the boundaries of the different regions
since the ground state exists without knowledge of the excited states. 
The contours in Fig.~\ref{fig2} for the excited state can only appear
in the regions with two or more stable bound states.  These contours
therefore must connect points of the boundaries between regions I and
II. They may cross continuously through region III precisely as the
ground state would cross through boundaries between regions I and II.
Similar contours exist within region III but we do not exhibit them in
this narrow strip where they are allowed.  The scaled three-body
energies are often substantially larger than the initial two-body
energies, although both arise from the same two-body interactions.

\begin{figure}[t!]
\includegraphics[scale=0.28,clip=true]{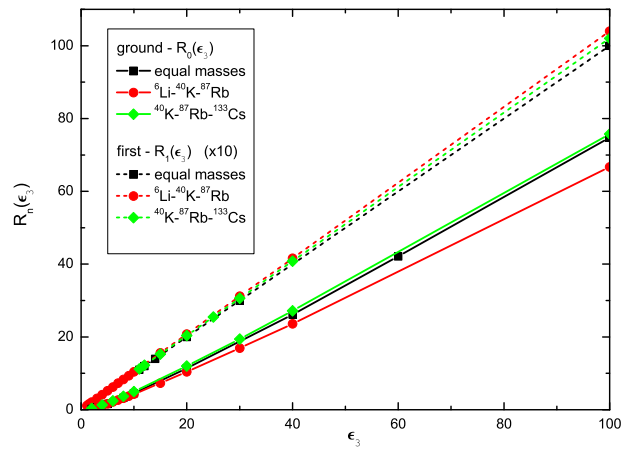}
\caption{$R_0$ and $R_1$ as function of $\epsilon_3$ 
  for three sets of mass ratios
  corresponding to $^{6}$Li$^{40}$K$^{87}$Rb and 
  $^{40}$K$^{87}$Rb$^{133}$Cs. The curves for the excited state
  show $10R_1(10\epsilon_3)$ in order to fit within the same 
  range as the results for the ground state.}
\label{fig3}
\end{figure}

\section{Parametrization.}
The universal functions, $F_n$, defined in Eq.~\eqref{e40} are not
easily found analytically. 
However, the contour diagrams in Fig.~\ref{fig2} suggest
a simple implicit dependence in terms of an extended Lam{\'e} curve
or {\it superellipse} \cite{lame1818}. Note that despite the 
log-log scale of Fig.~\ref{fig2}, the parametrization in terms
of Lam{\'e} curves is done with the energies on a linear scale. 
The three-body energies can be written
indirectly by {\it supercircles}, i.e.
\begin{equation}
 \epsilon_{ac}^{t_{n}} +  \epsilon_{bc}^{t_{n}} = R_n^{t_{n}} \;,
\label{e70}
\end{equation}
where the radius, $R_n$, and the power, $t_{n}$, are functions
of $\epsilon_3$ and both depend on the two mass ratios. The term
{\it supercircle} has been adopted since there is a {\it single} 
axis parameter, $R_n$. The
smallest value of $\epsilon_3$ is unity corresponding to the
two-body threshold of the $ab$ system used as the energy unit.

We choose sets of masses of current interest, and present the fitted
radius functions in Fig.~\ref{fig3} for both ground and first excited
states.  In $^{6}$Li$^{40}$K$^{87}$Rb a second excited state
exists whereas equal mass and $^{40}$K$^{87}$Rb$^{133}$Cs only have stable ground and
first excited states.  The parametrized results cannot be
distinguished from the computed curves in Fig.~\ref{fig2}.  The radius
functions turn out to be surprisingly simple, that is essentially
linear functions of $\epsilon_3$, and essentially independent of the
masses.  For the ground state we find a slight increase of slope with
increasing three-body energy.  Average estimates are 
\begin{eqnarray}
 R_0(\epsilon_{3}) \approx 0.74 \epsilon_{3}-2.5 \;\;,\;\;
R_1(\epsilon_{3}) &\approx& \epsilon_{3} \label{e80} \;.
\end{eqnarray}
The increasing functions reflect how the contours in Fig.~\ref{fig2}
move to larger two-body energies with increasing $\epsilon_{3}$.

\subsection{Symmetric systems.}
It is very satisfactory to observe this simple linear dependence which
implies that the three-body energy increases linearly with a kind of
average of the two two-body energy ratios.  It is satisfying also
because the symmetric system, where all particles are identical, has
this property where two- and three-body energies are proportional in
the universal limit.  
To approach this limit we assume that $a=b$, and Eqs.~\eqref{e70} and \eqref{e80} imply for
the ground states that $0.74 \epsilon_3 \approx 2.5 + \epsilon_{ac} 2^{1/t_0}$
When $a=b=c$, we get the known ratios
$\epsilon_3 \approx (2.5 + 2^{1/t_0})/0.74 = 16.52$. For the excited
state we get correspondingly $\epsilon_3 \approx 2^{1/t_1} =
1.267$. This is achieved with $t_0 \approx 0.30$ and $t_1 \approx 2.93$.

A system with particles $a$ and $c$ identical has been studied in Ref.~\cite{bel11}.  
Here $\epsilon_{bc} = 1$ and we can find
$\epsilon_{3}$ from Eqs.\eqref{e70} and \eqref{e80}. We get the
leading order terms for $\epsilon_{3}$ to be $\epsilon_{ac}/0.74$ and
$\epsilon_{ac}$ for large $\epsilon_{ac}$, respectively for ground and
excited states. For small $\epsilon_{ac}$ we find correspondingly that
$\epsilon_{3}$ approaches the constants $3.5/0.74$ and $1$.  These
leading order terms can be supplemented by further expansion to any
desired order. The main dependence of the parametrization is
consistent with the numerical results in \cite{bel11}.

\begin{figure}[t!]
\includegraphics[scale=0.28,clip=true]{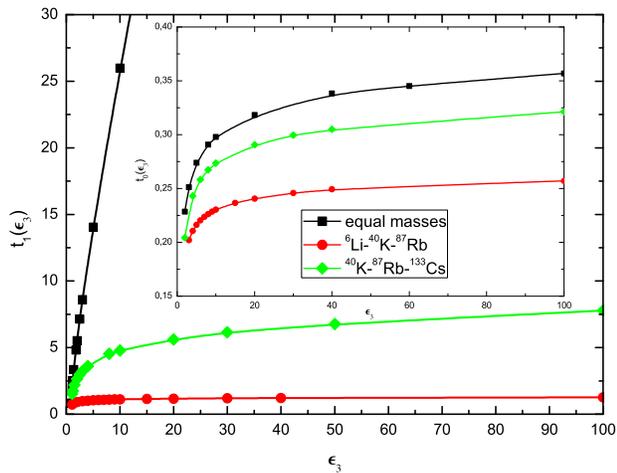}
\caption{Exponents, $t_1$ and $t_0$ (inset) in the {\it supercircle} fit
  for $^{6}$Li$^{40}$K$^{87}$Rb and 
  $^{40}$K$^{87}$Rb$^{133}$Cs as in Fig.~\ref{fig3}.}
\label{fig4}
\end{figure}

\subsection{The exponents.}
The exponents, $t_{n}$, are crucial to obtain the correct
curvature of the energy contours in Fig.~\ref{fig2}.  In
Fig.~\ref{fig4} we show the functions obtained
for the same sets of masses as in Fig.~\ref{fig3}.  These exponents
increase monotonously with $\epsilon_{3}$ from $0.2$ ($n=0$) and $1$
($n=1$) at the minimum value of $\epsilon_{3}=1$. The curves bend
over and eventually stop when the states reach a
two-body threshold and become unstable. In most cases this only
happens at large energies where the universal properties
are very unlikely.

The absolute sizes increase by about an order of magnitude from
ground to first excited state. The role of the exponents in
Eq.~\eqref{e70} is to describe the curvature of the contours in
Fig.~\ref{fig2}. Thus, large $t_n$ is necessary for strongly bending
curves. This explains the difference between ground and
first excited state, but also the overall increase with
$\epsilon_{3}$.  This is especially pronounced for the excited
states which are squeezed in between boundaries defined by stability
towards decay to bound two-body subsystems.

The contours connect boundaries with small $\epsilon_{ac}$ and large
$\epsilon_{bc}$ or vice versa. The end-points are when either $\epsilon_{bc}$ or
$\epsilon_{ac}$ is close to $\epsilon_{3}$. The slopes are from
Eq.~\eqref{e70} found to be, $\partial \epsilon_{ac}/ \partial
\epsilon_{bc} = -(\epsilon_{bc}/ \epsilon_{ac})^{t_n-1}$.  This means
that the contours in Fig.~\ref{fig2} first exhibit a small negative
slope (small $\epsilon_{bc}$) and then a quick bend to reach a large
negative slope (small $\epsilon_{ac}$).  All contours decrease from
left to right. The regions become narrower for higher two- and
three-body energies, and the contours must bend quickly.  The
implication is that $t_n$ must increase with $\epsilon_{3}$ and with
the number of excited states.

The allowed regions for a specific number of bound states are
indicated in Fig.~\ref{fig1}. The practical examples selected for
Figs.~\ref{fig3} and \ref{fig4} are found in the center, and in
regions II and III of the third quadrant. In Fig.~\ref{fig2} the
boundaries for two bound states are rather widely separated giving
rise to the smallest exponents in Fig.~\ref{fig4}. Moving towards more
bound states in Fig.~\ref{fig1} would widen this region and the narrow
strip with three bound states as well.  At some point one more bound
state appears but in the process all regions widened and the exponents
in the fits consequently decrease.  The other cases in
Figs.~\ref{fig3} and \ref{fig4} are both in region II of
Fig.~\ref{fig1} but the point for equal masses is further away from
the boundary to a region III than the point of asymmetric masses. This
is a general explanation of the hierarchy in the sizes of the
exponents.

The behavior of the exponents is also surprisingly simple for each
set of masses.  The relatively fast increase at small energies in
Fig.~\ref{fig4} slows down and both $t_0$ and $t_1$ approach constants
at large energy.  For the ground state, this can be accurately captured by
\begin{eqnarray}
t_0(\epsilon_{3}) \approx  \alpha_0 \frac{\epsilon_{3}^{p_0}+\beta_0}
{\epsilon_{3}^{p_0} + \gamma_0},\label{e90}
\end{eqnarray}
where $\alpha_0$ and $\alpha_1$ are the mass dependent constants
approached at large energy, see Fig.~\ref{fig4}. The parameters,
$(p_0,\alpha_0) \simeq (0.04-0.06,0.3-0.5)$, exhibits a small mass
dependence, whereas 
$(\beta_0,\gamma_0) \simeq -(0.93-0.95),-(0.82-0.87)$ 
are slightly negative but almost mass independent.  
We emphasize that stability requires $\epsilon_3 > 1$.
The value of $t_0$ for small $\epsilon_3 \approx 1$ is then in the
range of $t_0 \simeq 0.2-0.5$ as required to give the limiting value
of $\epsilon_3=16.52$. A similar parametrization for the exponent
corresponding to the excited state can be found.

\section{Conclusions and perspectives.}
The general spectrum for three interacting particles in the 
universal regime in two dimensions is currently unknown.
In the simplest case of three identical bosons, there is 
a lack of analytical results and one must resort to a 
numerical treatment at the moment. 
We numerically compute the universal three-body energies for short-range
interactions, different masses and interaction strengths.
The number of bound states varies from one and up depending on mass ratios
and two-body subsystem energies, with symmetric system having the 
fewest and the most for two heavy and a light particle. The
three-body ground state energies can be successfully parametrized by universal
functions that are so-called {\it supercircles} (powers different from two)
where the coordinates are the independent two-body energy ratios and the 
radius parameter is the three-body energy. The latter is an 
approximately linear function of the three-body energy, independent
of masses, while the powers of the coordinates are functions of 
both mass and three-body energy.
The simple threshold behavior for identical bosons
where the two three-body bound states are directly 
proportional is only valid for relatively small three-body energies.
Whether this indicates a smaller universal region has to be
investigated by either finite range models or radius computations.  Our
results can be used as
estimates of three-body energies and number of bound states, and as
a measure of the deviation from the universal zero-range limit.

\paragraph*{Acknowledgments} This work was partly support by funds
provided by FAPESP (Funda\c c\~ao de Amparo \`a Pesquisa do Estado
de S\~ao Paulo) and CNPq (Conselho Nacional de Desenvolvimento
Cient\'\i fico e Tecnol\'ogico ) of Brazil.

\end{document}